\documentclass{mn2e}
\usepackage{epsf}
\usepackage{times}
\newcommand{\etal}{{et al}\/.}
\begin{document}
\title[The X-ray jet in 3C\,66B]{{\it Chandra} observations of the
X-ray jet in 3C\,66B}
\author[M.J.~Hardcastle \etal]{M.J.\ Hardcastle, M.\ Birkinshaw and
D.M.\ Worrall\\
Department of Physics, University of Bristol, Tyndall Avenue,
Bristol BS8 1TL}
\maketitle
\begin{abstract}
Our {\it Chandra} observation of the FRI radio galaxy 3C\,66B has
resulted in the first detection of an X-ray counterpart to the
previously known radio, infra-red and optical jet. The X-ray jet is
detected up to 7 arcsec from the core and has a steep X-ray spectrum,
$\alpha \approx 1.3 \pm 0.1$. The overall X-ray flux density and
spectrum of the jet are consistent with a synchrotron origin for the
X-ray emission. However, the inner knot in the jet has a higher
ratio of X-ray to radio emission than the others. This suggests that
either two distinct emission processes are present or that differences
in the acceleration mechanism are required; there may be a
contribution to the inner knot's emission from an inverse-Compton
process or it may be the site of an early strong shock in the jet. The
peak of the brightest radio and X-ray knot is significantly closer to
the nucleus in the X-ray than in the radio, which may suggest that the
knots are privileged sites for high-energy particle
acceleration. 3C\,66B's jet is similar both in overall spectral shape
and in structural detail to those in more nearby sources such as M87
and Centaurus A.
\end{abstract}
\begin{keywords}
galaxies: active -- X-rays: galaxies -- galaxies: individual: 3C\,66B
-- radiation mechanisms: non-thermal
\end{keywords}

\section{Introduction}

In the first year of its operation, the {\it Chandra X-ray Observatory}
has vastly increased the number of known X-ray counterparts to radio
features of extragalactic radio sources. Several X-ray detections of
hotspots in classical double sources (e.g.\ Harris, Carilli \& Perley
1994; Harris, Leighly \& Leahy 1998) had already been made with {\it
ROSAT}, but {\it Chandra} has detected several new objects (3C\,295,
Harris \etal\ 2000; 3C\,123, Hardcastle, Birkinshaw \& Worrall 2001;
3C\,263, Hardcastle \etal\ in preparation) as well as giving new
spectral information on existing detections (Wilson, Young \& Shopbell
2000, 2001). However, the most striking result has been the detection
of a number of new X-ray jets. Individual knots in the jets of some
prominent objects had already been detected in the X-ray in a few
cases (e.g. M87, Harris \etal\ 1997, Neumann \etal\ 1997; Centaurus A,
D\"obereiner \etal\ 1996, Turner \etal\ 1997; 3C\,273, R\"oser \etal\
2000) but the superior sensitivity and resolution of {\it Chandra} has
allowed the imaging of a number of continuous features which meet the
Bridle \& Perley (1984) criteria for classification as a jet. The
early detections were of powerful FRII objects (PKS 0637$-$752,
Schwartz \etal\ 2000, Chartas \etal\ 2000; Pictor A, Wilson \etal\
2000) but it is becoming clear that X-ray jets are also common in
nearby FRI sources. This was unexpected, since optical FRI jets are
rare and since inverse-Compton processes in FRIs might be expected to
produce only very faint emission. In a previous paper (Worrall,
Birkinshaw \& Hardcastle 2001) we reported on the first two new FRI
X-ray jets to be discovered with {\it Chandra}, based on short ($\sim
10$ ks) {\it Chandra} observations. In this paper we present a deep
observation of a third new X-ray jet, in the nearby radio galaxy
3C\,66B.

3C\,66B is a low-luminosity FRI radio galaxy with a redshift of
0.0215, hosted by the elliptical galaxy UGC 1841, which lies in a
small group on the edge of the cluster Abell 347, part of the
Perseus-Pisces supercluster. In the radio, its bright jet was one of
the first to be discovered (Northover 1973) and it has subsequently
been well studied (e.g.\ Leahy, J\"agers \& Pooley 1986). Detailed
comparisons have been made between the jet and the weaker counterjet
(Hardcastle \etal\ 1996, hereafter H96). An optical counterpart to the
jet was discovered by Butcher, van Breugel \& Miley (1980), and
identified from its polarization properties as synchrotron radiation
(Fraix-Burnet \etal\ 1989). The optical jet has since been observed at
a number of wavelengths with the {\it Hubble Space Telescope} ({\it
HST}) FOC and STIS (Macchetto \etal\ 1991; Sparks \etal\ in
preparation), and the optical structure has been shown to agree well
with the radio (Jackson \etal\ 1993). Most recently, Tansley \etal\
(2000) have used {\it ISO} to detect a mid-infrared jet whose
properties are consistent with those found at other
wavebands. Giovannini \etal\ (2001) report on global VLBI observations
which show that the high jet to counterjet ratio persists to sub-pc
scales. The counterjet, if present, is below the noise limit for most
of the length of the pc-scale jet, although they do detect a
counterjet component very close to the VLBI core, perhaps implying jet
velocity structure or variations in velocity along the jet. They
suggest that the source lies at about 45$^\circ$ to the line of sight,
consistent with earlier inferences from the kpc-scale jet asymmetries
(H96).

In the X-ray, some extended emission associated with the source, as
well as compact nuclear emission, was detected with {\it Einstein}
(Maccagni \& Tarenghi 1981), and also appears in {\it ROSAT} Position
Sensitive Proportional Counter (PSPC) All-Sky
Survey data (Miller \etal\ 1999); however, the source appeared
unresolved to the {\it ROSAT} High Resolution Imager (HRI; Hardcastle
\& Worrall 1999). Neither observatory had the resolution or
sensitivity to detect a possible counterpart to the radio and optical
jet, which is bright only out to about 7 arcsec from the nucleus. Our
observation with {\it Chandra} was intended to search for the X-ray
jet. In a future paper we expect to present results from our approved
{\it XMM-Newton} observations of 3C\,66B's gaseous environment.

Throughout this paper we assume $H_0 = 50$ km s$^{-1}$ Mpc$^{-1}$. One
arcsec corresponds to 0.61 kpc in this cosmology at the redshift of
3C\,66B. Spectral index $\alpha$ is defined in the sense $S \propto
\nu^{-\alpha}$.

\section{Observations and data processing}

\begin{figure*}
\epsfxsize 12cm
\epsfbox{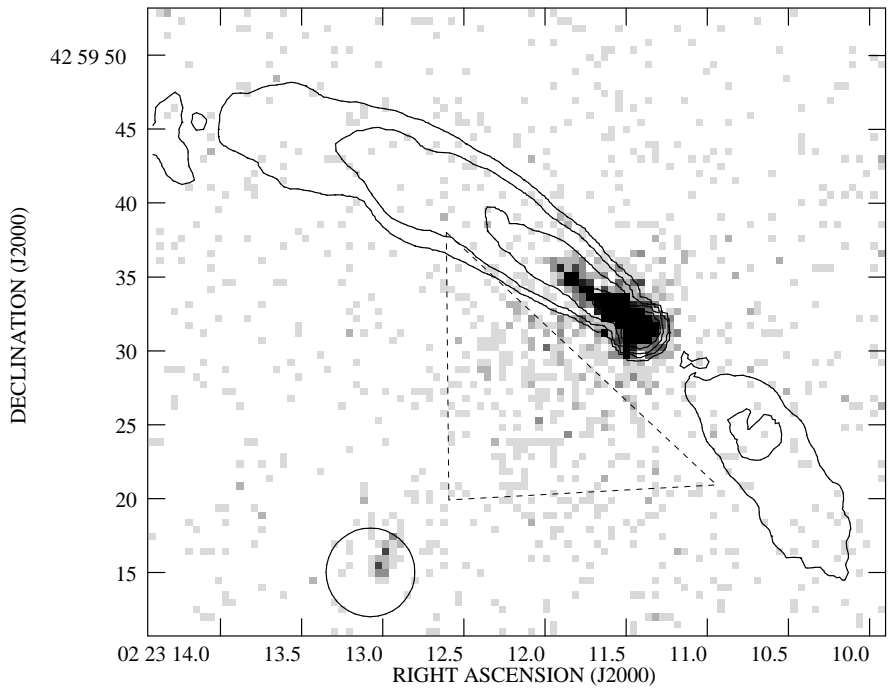}
\caption{Exposure-corrected 0.5--7.0 keV {\it Chandra} image of the
central region of 3C\,66B. Linear greyscale: black is $4 \times
10^{-7}$ photons cm$^{-2}$ s$^{-1}$ between 0.5 and 7.0 keV per
standard Chandra pixel (0.492 arcsec on a side). Superposed are
contours at $0.2 \times (1,4,16,64,256)$ mJy beam$^{-1}$ from an
8.2-GHz Very Large Array (VLA) map with 1.25-arcsec resolution (H96),
showing the jet and counterjet. The circle to the SE shows the
position of 3C\,66B's companion galaxy. The dotted lines show the
region from which the spectrum of the inner extended emission was
extracted.}
\label{image}
\end{figure*}

We observed 3C\,66B for 43968 s with {\it Chandra} on 2000 November
20th. Because it was possible (based on {\it ROSAT} fluxes) that the
count rate from the nucleus would give rise to pileup, we chose to
read out only 400 rows from each of the 6 ACIS-S CCDs, thus reducing
the frame time. 3C\,66B was positioned in the centre of the readout
region, near the aim point of the back-illuminated S3 chip. We
followed the {\sc ciao} `science threads' using {\sc ciao} version
2.0.2 to generate a new Level 2 events file with the 0.5-pixel
randomization removed and containing events with the standard set of
grades. The observation was affected by the well-known high-background
problem, and after filtering to remove the most strongly affected
intervals we were left with an effective exposure of 32762
s. Fig. \ref{image} shows the {\it Chandra} image produced from these
data. Diffuse emission on scales comparable to the optical galaxy, a
nuclear component and the jet are all detected in X-rays, and these
are discussed further in the following sections. In each case, spectra were
extracted using {\sc ciao}, with the best available responses being
constructed for each extraction region, and analysed using {\sc
xspec}. Spectra were binned such that every bin had $>20$ net counts,
and fits were restricted to the energy range 0.5--7.0 keV.

\section{Extended emission}

Some extended emission can clearly be seen in Fig.\ \ref{image}, lying
predominantly to the SE of the nucleus. Emission from the southeastern
companion galaxy of UGC 1841 is also detected, with about 50 net
counts. Radial profile analysis and examination of smoothed or binned
images show that 3C\,66B lies near the centre of a relatively smooth,
low-surface-brightness distribution of X-ray-emitting gas which
extends out to at least 40 arcsec from the core and contributes $\sim
500$ 0.5--7.0 keV counts to the image. Because the counts in much of
this region are dominated by background, and because the extended
emission very likely fills our readout region, it is difficult to
obtain a useful spectrum of this large-scale emission. However, we
have measured the spectrum of the clump of gas to the SE of the
source, which contains $160 \pm 30$ 0.5--7.0 keV counts using an
off-source background region (a source-centred annulus between 55 and
60 arcsec from the core). It is well fitted ($\chi^2/n = 3.2/7$, where
$n$ denotes degrees of freedom) with a {\sc mekal} model
with an assumed 0.5 cosmic abundance and $kT = 0.78 \pm 0.04$ keV
(errors are $1\sigma$ for one interesting parameter throughout), using
Galactic absorption ($N_{\rm H} = 6.84 \times 10^{20}$ cm$^{-2}$;
Stark \etal\ 1992). This temperature is similar to that seen in
extended gas close to the host galaxies of other FRI sources (Worrall
\& Birkinshaw 2000, Worrall \etal\ 2001).

The large-scale bending of the jets in 3C\,66B, seen in Fig.\
\ref{image}, is plausibly caused by the motion of the host galaxy through the
group-scale gas in a northwesterly direction. If this is the case, it
is interesting that the high-surface-brightness extended emission lies
behind the galaxy, to the SE. This may be evidence for ongoing
ram-pressure stripping of hot gas associated with UGC 1841.

The small number of counts from the companion galaxy prevents us from
analysing its spectrum in detail, but its emission appears
significantly harder than that of the nearby extended gas; the
best-fitting $kT$, assuming Galactic absorption, is 2.7 keV, and is
greater than 1.0 keV at the 99\% confidence level. The X-ray emission
appears to be slightly off-centre, in the direction of 3C\,66B. The
luminosity of the detected emission is about $2 \times 10^{40}$ ergs
s$^{-1}$, so it is plausible that it originates in a population of
X-ray binaries.

\section{The nucleus}

The nucleus of 3C\,66B produced $1259 \pm 36$ counts between 0.5 and
7.0 keV, using an extraction circle with radius 3 {\it Chandra} pixels
(1.5 arcsec) to avoid contamination with jet and thermal emission, and
taking background from an off-source region (a 30-arcsec circle
centred 48 arcsec to the SE of 3C\,66B). Its spectrum is adequately
fitted ($\chi^2/n = 62.0/48$) with a simple power-law model and
Galactic absorption, in which case the energy index $\alpha = 1.14 \pm
0.05$. The fit is improved ($\chi^2/n = 57.9/47$) if the Galactic
$N_{\rm H}$ is allowed to vary; the improvement is significant at the
90 per cent level on an F-test and the best-fitting value of $N_{\rm
H}$ is $(12 \pm 2) \times 10^{20}$ cm$^{-2}$, with $\alpha =
1.35_{-0.12}^{+0.13}$. A similar improvement ($\chi^2/n = 58.0/47$) is
obtained if the local absorption column is fixed at the Galactic value
and the excess absorption is intrinsic to the nucleus of UGC 1841, in
which case the intrinsic column is $(5 \pm 2) \times 10^{20}$
cm$^{-2}$.  Finally, we can obtain slightly improved fits ($\chi^2/n =
53.7/46$; the improvement is again significant at the 90 per cent
level) if there is a contribution to the core spectrum at the $\sim 6$
per cent level (in terms of total flux in the 0.5--7.0 keV band) from
hot gas with $kT = 0.56_{-0.10}^{+0.07}$ keV, fixing the absorption
column for both thermal and power-law components at the Galactic value
and the abundance at 0.5 cosmic. This temperature is similar to that
seen in the off-nuclear extended emission and the addition of this
component has the effect of flattening the best-fitting core spectrum,
$\alpha = 1.03_{-0.05}^{+0.07}$. The normalization of the thermal
component implies a gas density somewhat higher than that in the
extended emission discussed above, but this is consistent with
our expectations, and so we adopt this model. The low intrinsic absorption
and steep spectrum of the core are consistent with the results for
other {\it Chandra}-observed FRI radio galaxies in Worrall \etal\
(2001).

The unabsorbed 0.5--4.0 keV luminosity of the non-thermal component of
the core in our adopted model is $3 \times 10^{41}$ ergs s$^{-1}$.
The 1-keV flux density is $30 \pm 2$ nJy (values based on other models
are similar). This is substantially lower than the value of $141 \pm
33$ nJy that we quoted, based on the {\it ROSAT} HRI data, in
Hardcastle \& Worrall (1999). But when we take into account the fact
that the {\it ROSAT} data point would have included a contribution
from the jet and the extended emission in the 1.5-arcmin source circle
used to determine the HRI count rate, the difference becomes less
significant. The {\it ROSAT} data still predict around 40 per cent
more counts in the central regions of the source than are observed
with {\it Chandra}, but the {\it ROSAT} count rates have uncertainties
of around 25 per cent, so that the difference is at less than the
$2\sigma$ level. We cannot conclude that there is significant evidence
for X-ray nuclear variability.

The centroid of the X-ray nucleus is offset by about 0.22 arcsec to
the east of the radio nucleus, whose position is known to high
accuracy. Since this discrepancy is within the known
uncertainties in {\it Chandra} astrometry, we have assumed that the
radio and X-ray nuclei coincide, and have shifted the {\it
Chandra} data accordingly in what follows.

\section{The jet}

\begin{figure*}
\epsfxsize 12cm
\epsfbox{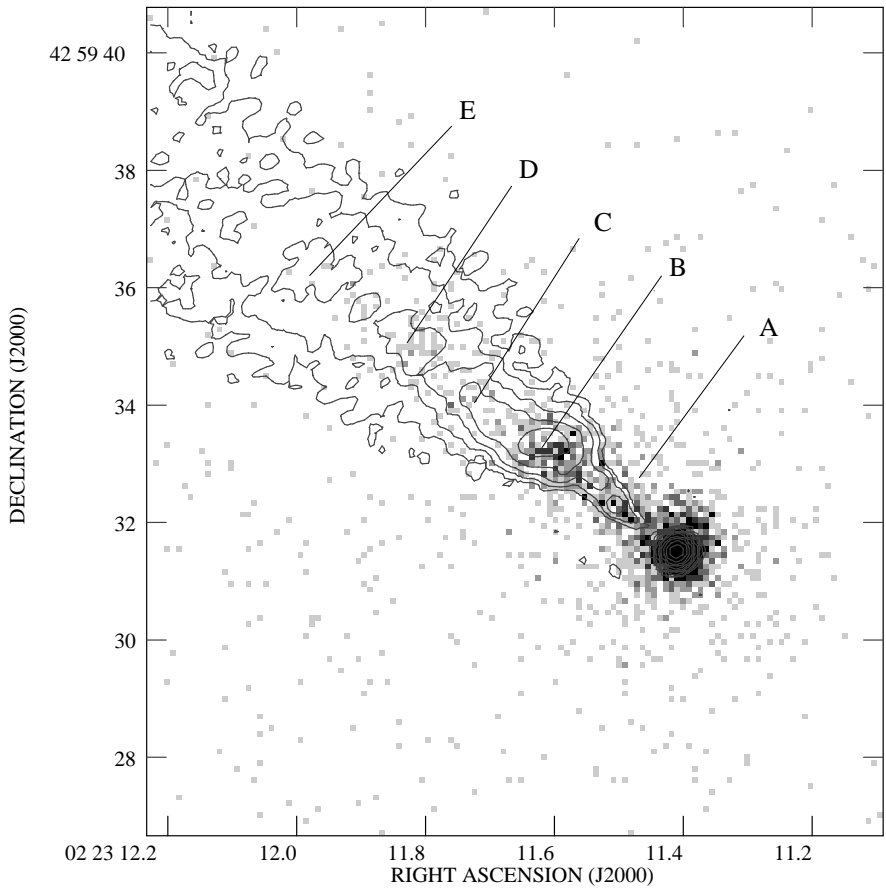}
\caption{The jet and core of 3C\,66B. Each pixel is 0.0984 arcsec on a
side; black is 5 counts pixel$^{-1}$ between 0.5 and 7.0
keV. Superposed are contours at $0.1 (\times 1,2,4,8,16,\dots)$ mJy
beam$^{-1}$ from an 8.2-GHz VLA map with 0.25-arcsec resolution
(H96). The labels show the names of the knots in the notation used by
Fraix-Burnet \etal\ (1989) and H96.}
\label{hijet}
\end{figure*}

Fig.\ \ref{hijet} shows a high-resolution image of the jet overlaid
with contours from the 8.2-GHz VLA map of H96.
X-ray emission from the jet is seen out to 7 arcsec from
the core, or somewhere between knots D and E in the
notation of Fraix-Burnet \etal\ (1989), and may continue to larger
distances at low significance. The X-ray emitting region is thus a
good match to the optical synchrotron jet as seen by the {\it HST}.
As with the optical emission, it seems that the X-rays are coming from
the jet as a whole rather than just from individual knots; there is
clearly emission from the inter-knot regions between knots B, C and D.

The jet (using a polygonal extraction region enclosing the jet
emission seen on the high-resolution map of Fig.\ \ref{hijet})
contains $500 \pm 23$ counts (0.5--7.0 keV), using the same off-source
background region as for the nucleus. The spectrum is adequately
fitted ($\chi^2/n = 20.5/20$) with Galactic absorption and a single
power law of $\alpha = 1.31 \pm 0.09$, which gives a 1-keV jet flux
density of $14 \pm 1$ nJy. The fit is not significantly improved
($\chi^2/n = 19.6/19$) if the Galactic $N_{\rm H}$ is allowed to vary;
the best-fitting $N_{\rm H}$ ($(11 \pm 5) \times 10^{20}$ cm$^{-2}$)
is consistent at the $1\sigma$ level with the best-fitting value
obtained for the core, but also with a Galactic absorption
column. Again, the best fits ($\chi^2/n = 16.0/18$, an improvement
only significant at the 90 per cent level on an F-test) are obtained
with a weak contribution from thermal gas with an emission measure
similar to that obtained in the fit to the extended emission and with
$kT = 0.50^{+0.08}_{-0.10}$ keV, in which case $\alpha =
1.13_{-0.13}^{+0.14}$. The net flux density of the non-thermal
component in this model is $12 \pm 1$ nJy. In order to simplify
comparisons later in the paper between the jet spectrum and that of
individual knots, we will use the simple power-law model with Galactic
absorption in what follows, noting that both fits are consistent at
the $1\sigma$ level with a power-law index $\sim 1.2$ and a 1-keV flux
density of 13 nJy.

\begin{figure}
\epsfxsize 8.5cm
\epsfbox{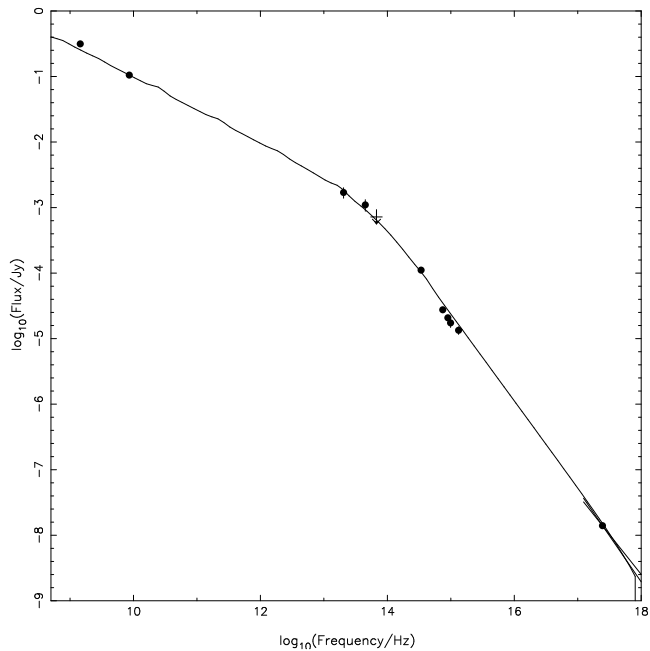}
\caption{The radio-to-X-ray spectrum of the jet of 3C\,66B. Data
points, with the exception of the X-ray point, are taken from Tansley
\etal\ (2000), with the {\it HST} points corrected for Galactic
reddening as described in the text. The X-ray flux density is derived
from the simple power-law fit to the {\it Chandra} data and the `bow
tie' illustrates the $1\sigma$ range of spectral index in this
model. The solid line shows the broken power-law synchrotron model
discussed in the text.}
\label{netspec}
\end{figure}

We can compare the integrated flux density from the jet to that seen in
other wavebands. Tansley \etal\ (2000) tabulated flux densities for
the region of the jet out to knot E (8 arcsec from the core). In Fig.\
\ref{netspec} we plot the rest-frame spectrum of this region of the jet. Note
that we have corrected the {\it HST} flux densities taken from Jackson
\etal\ (1993) both for the larger aperture (as in Tansley \etal) and
for the effects of Galactic reddening, using the $A_B$ of 0.40 quoted
by Burstein \& Heiles (1984). It can be seen that when this is done
the net radio-to-X-ray spectrum can be modelled as synchrotron
radiation from electrons with a broken power-law energy distribution,
with a flat spectrum at low frequencies, $\alpha = 0.5$, steepening in
the infra-red to $\alpha \approx 1.35$ and retaining this value out to
the X-ray. The spectral index measured from the X-ray data is
consistent with this picture (within $1\sigma$, if the simple
power-law model is adopted), and the fact that the {\it HST}
data points lie slightly below the line could be attributed to
uncertainties in the value of Galactic reddening (the best-fit values
of $N_{\rm H}$ derived above would imply higher values for $A_{\rm B}$
than the value given by Burstein \& Heiles, for example). Since the radio and
optical emission of the jet is known to be synchrotron, we might infer
from this that the X-ray emission is also all synchrotron in origin.

\begin{table*}
\caption{Radio, optical and X-ray parameters of knots A and B}
\label{knots}
\begin{tabular}{lrrrrrrrr}
\hline
Knot&8.2-GHz flux&Size&Equipartition&$\alpha_R$&1-keV flux&$\alpha_X$&$\alpha_{OX}$&$\alpha_{RX}$\\
&density (mJy)&(arcsec)&B-field (nT)&&density (nJy)\\
\hline
A&2.72&$1.7 \times 0.2$&10&$0.75\pm0.05$&$4.0 \pm 0.3$&$0.97 \pm 0.34$&$0.89\pm
0.04$&$0.78 \pm 0.03$\\
B&25.1&$1.1 \times 0.4$&12&$0.60\pm0.02$&$6.1 \pm 0.4$&$1.17\pm 0.14$&$1.27 \pm
0.01$&$0.89 \pm 0.01$\\
\hline
\end{tabular}
\vskip 2pt
\begin{minipage}{15.6cm}
Radio spectral indices are measured from our MERLIN 1.5-GHz and VLA
8.2-GHz images. X-ray spectra and flux densities use the spectral
model for the jet described in the text, with Galactic absorption
(unabsorbed flux densities are quoted). Equipartition field strengths
are calculated neglecting the effects of Doppler boosting and
projection, and assume a low-frequency spectral index of 0.5 and a
low-energy cutoff corresponding to $\gamma = 1000$, with no
contribution to the energy density from non-radiating particles such
as relativistic protons. Optical flux densities used to calculate spectral
indices are measured from the {\it HST} STIS images.
\end{minipage}
\end{table*}

\begin{figure}
\epsfxsize 8.5cm
\epsfbox{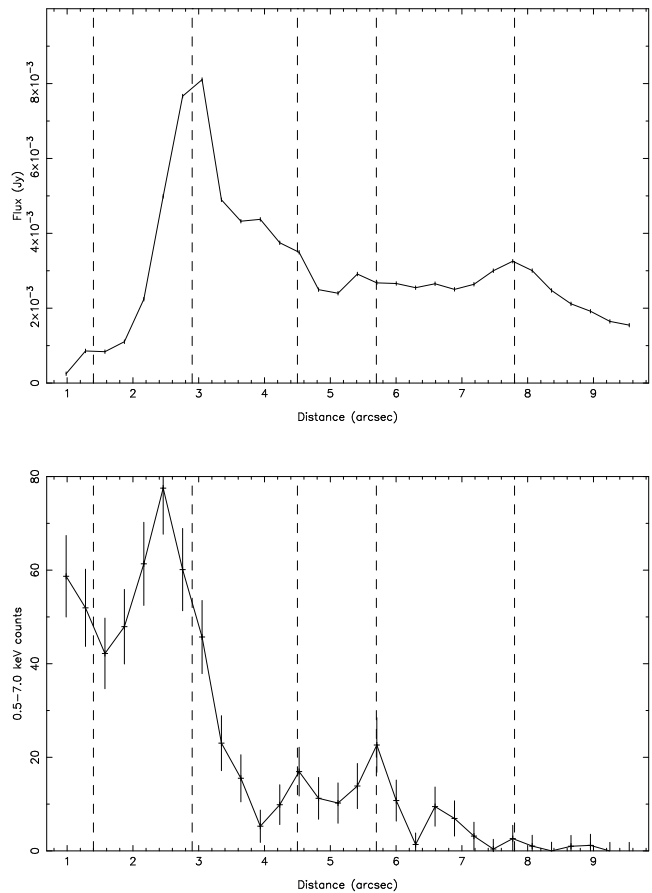}
\caption{Profiles of the jet in the radio (upper panel) and X-ray
(lower panel). Profiles were made using the maps of Fig.\ \ref{hijet},
extracting all the emission from a series of rectangles of 0.3 arcsec
by 2.0 arcsec (long axis perpendicular to the jet axis). No background
subtraction has been carried out. The innermost bin of the X-ray
profile is dominated by emission from the X-ray nucleus of
3C\,66B. Dashed lines show the approximate positions of knots A, B, C,
D and E (in order of increasing distance from the nucleus). Errors on
the points on the X-ray profile are $1\sigma$; note that the X-ray
data points are not independent, since the sampling along the jet axis
is smaller than the resolution of {\it Chandra}. Errors on the radio
data points are $1\sigma$ based on the off-source noise, and so are
lower limits on the actual errors.}
\label{profile}
\end{figure}

\begin{figure}
\epsfxsize 8.5cm
\epsfbox{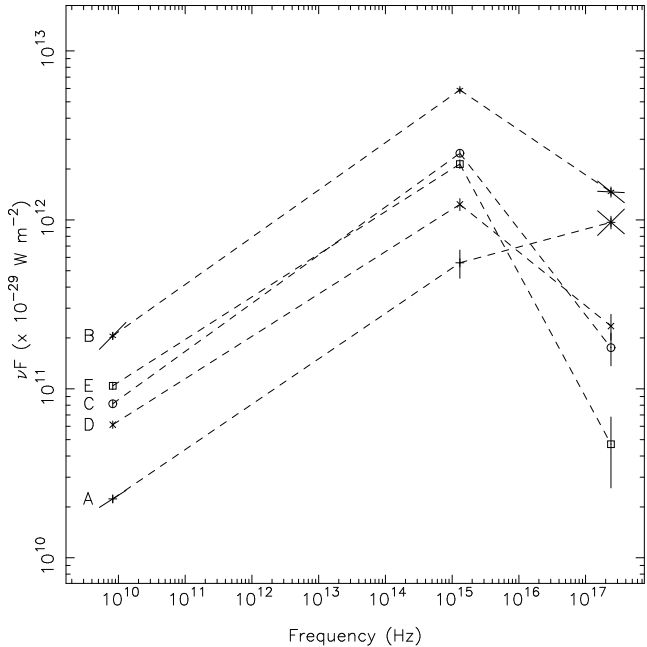}
\caption{$\nu F_\nu$ plot of the flux densities of the five
radio-optical-X-ray knots in 3C\,66B. These were measured from maps
generated with the same pixel scale and aligned at the nucleus, using
identical polygonal extraction regions, defined on the X-ray map,
within {\sc miriad} (except for the UV flux of knot A, which was
obtained using a smaller region to minimise systematic errors from
background subtraction). The solid bars for knots A and B show the
known radio and X-ray spectral indices (Table \ref{knots}). The dotted
lines are intended to guide the eye only, and do not represent a
spectral fit. The maps used are the 8.2-GHz image of H96,
the archival {\it HST} STIS image (Sparks \etal\ in preparation) and the {\it
Chandra} image. For the {\it Chandra} and {\it STIS} images adjacent
background regions were used to carry out background subtraction. The
STIS flux densities, converted from count rates using factors
calculated with the {\sc iraf} task {\sc calcphot} assuming an optical
spectral index of 1.3, have been corrected for an estimated 0.85
magnitudes of Galactic reddening at 2300 \AA. The conversion of {\it
Chandra} counts to flux densities is based on the simple power-law
fit to the integrated spectrum of the jet. Errors
plotted are statistical for the UV and X-ray images, and the standard
estimated 5 per cent flux calibration errors for the radio data
points; additional systematic uncertainties on the conversion of STIS
and {\it Chandra} counts to flux densities will not affect the
relative ordering of points on the plot.}
\label{knotflux}
\end{figure}

However, a more complicated picture is necessary when we look at the
X-ray jet in more detail. Fig.\ \ref{hijet} shows that the
radio-to-X-ray ratio is by no means constant along the jet. The
innermost regions of the jet, within 2 arcsec of the core, are
responsible for a large fraction of the jet X-ray emission, but a
negligible fraction of the total jet radio flux. The ratio of X-ray
counts to radio flux in this region (traditionally denoted `knot A',
though in fact it describes the whole of the inner jet before the
flare point, consisting of two knots and some extended emission) is a
factor $\sim 6$ higher than in the brightest radio knot, knot B, at
about 3 arcsec from the nucleus. This can be seen in Fig.\
\ref{profile}, which shows the X-ray and radio fluxes as a function of
distance along the jet. The difference arises somewhere between the
optical and X-ray regimes, as illustrated in Fig.\ \ref{knotflux},
which shows the flux density of each knot in the radio, ultraviolet
and X-ray bands. All the knots have similar radio-to-optical spectra
and show spectral steepening between the optical and X-ray, but knot
A's optical-to-X-ray spectrum is much the flattest. This has the
effect that knot A, though the faintest radio and optical knot, is the
second brightest X-ray feature in the source. The radio spectral index
(between 1.5 and 8.2 GHz) of knot A is also different from the other
knots; using matched regions on the VLA and our unpublished L-band
MERLIN maps, we estimate that $\alpha_{1.5}^{8.2}$ is about 0.75 for
knot A and typically 0.6 for the other knots. However, the X-ray
spectrum of knot A, using an adjacent background region, is
indistinguishable from that of the rest of the jet, within the errors,
at $\alpha = 1.0\pm 0.3$ (assuming the Galactic absorption column and
fitting a single power law). Table \ref{knots} compares the parameters
of knots A and B. Fig.\ \ref{knotflux} also shows that knots C and E
have a steeper optical-to-X-ray spectrum than knots B and D, although
the errors are large.

The offset of $0.35 \pm 0.1$ arcsec between the X-ray and radio peaks
of knot B (Fig.\ \ref{profile}), with the X-ray peak being closer to
the core, is a notable feature of the jet.  There is no offset of this
magnitude between the optical and radio profiles of the jet either here
or at any other point along its length (cf. Jackson \etal\ 1993), so
this is an effect which only appears at high energies. In contrast,
the X-ray knots C and D seem very well aligned with their radio counterparts.

No X-ray emission is detected from the counterjet in 3C\,66B. The
$3\sigma$ upper limit on emission from the counterjet in a region
similar to the region in which jet emission is seen is $\sim 20$ net
counts, or about 0.5 nJy, an X-ray jet-counterjet ratio of
$\ga 25$. The radio jet-counterjet ratio in this region is $\sim
20$, which is consistent; these data indicate that the counterjet
does not have a flatter radio-to-X-ray spectrum than the jet.

\section{Emission processes}

What process is responsible for the X-ray emission from the jet of
3C\,66B?

For knots B--E a simple synchrotron explanation still seems viable; in
fact, removing the X-ray flux due to knot A from the integrated flux
densities plotted in Fig. \ref{netspec} would improve the agreement
between the X-ray flux and the extrapolation from the optical data.
The differences between the radio-optical-X-ray spectra of the
individual knots (Fig.\ \ref{knotflux}) can be accommodated without
too much difficulty in a model in which the electron spectra and/or
magnetic field strengths are slightly different in the different
knots.

There are two obvious difficulties faced by a synchrotron model
for the X-rays. Firstly, there is the necessity to produce the
inferred electron spectrum, which is not a conventional
continuous-injection spectrum, since it breaks by $\sim 0.85$ in
spectral index, nor a conventional aged synchrotron spectrum, where
the curvature is due to a high-energy cutoff so that spectral index
increases with frequency after some critical frequency. Secondly,
there is the extremely short synchrotron loss timescale for
X-ray-emitting electrons, approximately 30 years for electrons
radiating at 1 keV in knot B in 3C\,66B if we assume the magnetic
field strength is close to the equipartition value of 12 nT (Table
\ref{knots}). This loss timescale rules out the possibility that the
acceleration of the synchrotron-emitting particles takes place in a
single region of the jet, for example in knot A or in the nucleus (a
possibility which was in any case only remotely viable given the
optical emission; H96). Even if the magnetic field strength is
extremely low, inverse-Compton losses due to scattering of microwave
background photons and starlight
limit the lifetime of the electrons emitting in the X-ray. This
considerably constrains so-called `low-loss channel' models in which
the electrons are transported in a region of low magnetic field
strength to the required places in the jet. If we require that the
critical frequency $\nu_c$ of electrons, after they have left the
low-loss channel and started to emit substantial amounts of
synchrotron radiation, be above the X-ray band, then (e.g.\ Leahy 1991)
\[
\nu_{\rm c} = {{C B_{\rm h}}\over{\left(B_{\rm l}^2 + B_{\rm p}^2\right)^2 t^2}} \ga
10^9{\rm\ GHz}
\]
where $B_{\rm h}$ is the magnetic field strength in the emitting
regions, $B_{\rm l}$ is the magnetic field strength in the low-loss
channel, $B_{\rm p}$ is the effective magnetic field strength due to
inverse-Compton losses ($B_{\rm p}^2 = 2\mu_0 U_{\rm p}$, where
$U_{\rm p}$ is the energy density in photons), and $t$ is the time
spent in the low-loss channel; $C$ is a constant with value, in
appropriate units, $2.5 \times 10^{15}$ nT$^3$ yr$^2$ GHz.  Even in
the limit that $B_{\rm l}$ is zero, this means that the time $t$ spent
in the channel must be $< 1.6 \times 10^3 \sqrt{B_{\rm h}}/B_{\rm
p}^2$ yr (where $B_{\rm h}$ and $B_{\rm p}$ are in nT), while the
light travel time to the end of the jet is $>1.5 \times 10^4$ yr (and
plausible electron travel times to this location are probably a factor
of a few higher). For $B_{\rm p} \approx 1.6$ nT, the value used by
H96 (the photon energy density is assumed to be dominated by
starlight), these times cannot be reconciled unless $B_{\rm h} \ga
600$ nT, two orders of magnitude above equipartition values near the
end of the jet. Since the effective energy density in CMB photons in
the jet frame increases with the bulk Lorentz factor (see below), it
is not possible to evade this constraint by requiring the flow through
the low-loss channel to be highly relativistic. We therefore regard
such models as implausible. Instead, there must be regions at which
acceleration takes place distributed throughout the jet -- `{\it in
situ} acceleration' -- and these must produce similar electron energy
spectra at all points beyond knot A. {\it Chandra\/}'s spatial
resolution does not allow us to distinguish between a model in which
there are a number of discrete acceleration regions along the jet
spine and one in which acceleration takes place everywhere in the jet,
but the lack of clear radio-optical spectral-index gradients (Jackson
\etal\ 1993) suggests that the latter is the case. However, the offset
between the X-ray and radio peaks of knot B might support a model in
which the bright radio knots are privileged sites for high-energy
particle acceleration; in this model the shock is located close to the
X-ray peak, and the offset is caused by rapid synchrotron losses of
the X-ray emitting particles in the high magnetic field strength of
knot B as they stream away from the shock.  The more strongly peaked
X-ray emission profile (Fig.\ \ref{profile}) may be additional
evidence to support this picture.

There is also no clear reason why a synchrotron model should not
describe knot A, but we would like to be able to understand why
synchrotron emission there is so different from the rest of the
jet. Because the spectral break in the integrated jet spectrum lies in
the infra-red, it is hard to construct a model in which simple
differences in physical conditions, such as magnetic field strength,
are responsible for the differences in the ultraviolet-to-X-ray
spectrum in knot A; they would be expected to affect the STIS flux of
the knot too. So, if the knot A X-rays are synchrotron emission, the
high-energy electron spectrum must be quite different there. If the
difference in the low-frequency radio spectral indices of knot A and
the rest of the jet is significant, the electron energy spectrum of
knot A must be different at all energies represented by our
observations. Since the high-energy electrons are accelerated {\it in
situ}, this implies in turn that the acceleration process must be
different in some way in this part of the jet. [Since the
equipartition magnetic field strength in knot A is similar to that in
knot B, as shown in Table \ref{knots} (the much lower radio flux
density is offset by the smaller physical size of the region) the
electron synchrotron ages are similar to those quoted above, and so it
is difficult to avoid the necessity of {\it in situ} acceleration in
knot A.] If we believe that the inner jet, represented by knot A, is
physically different from the rest of the jet, then differences in the
acceleration process there would perhaps not be surprising; one
possible model, discussed in H96, is that the knot A region represents
a supersonic, relativistic jet, like those in FRII sources, which then
disrupts and becomes trans-sonic and turbulent at knot B. Since the
location of acceleration of the particles responsible for emission
from FRII-type jets is far from clear, this model makes few concrete
predictions. Alternatively, knot A may just be the first in a series
of oblique shocks that take place as the jet decelerates. In this
case, the knot A shock or shocks would be the strongest, and could be
the most efficient at producing a population of high-energy particles;
the X-ray excess in knot A could be explained as synchrotron emission
from this newly accelerated electron population.

What are the alternatives to a synchrotron model for the excess
emission in knot A? Thermal emission seems implausible; if the jet is
intrinsically two-sided, there is no reason to expect to see thermal
emission only on the side oriented towards us. The most promising
alternative to synchrotron is inverse-Compton (IC) scattering by the
relativistic electrons in the jet of some population of lower-energy
seed photons, a process which should produce X-ray emission with a
spectral index similar to that of the radio data, $\alpha \sim
0.75$. The X-ray spectral index of knot A is somewhat steeper than
this, so that an IC origin for its emission is
contraindicated. However, we cannot rule out the possibility at a high
confidence level given the uncertainties on our X-ray spectrum.

There are four possible seed photon populations for IC emission; in
increasing order of their likely lab-frame energy density, these are
the microwave background (CMB), synchrotron photons, starlight, and
photons from the beamed nucleus of 3C\,66B (BL Lac light). Of these,
the first three are too weak to produce significant X-ray emission
unless the magnetic field strength in knot A is about 2 orders of
magnitude below the equipartition/minimum energy level. As Tavecchio
\etal\ (2000) and Celotti, Ghisellini \& Chiaberge (2001) have pointed
out, the effective energy density of CMB photons in the jet frame is
increased significantly if the jet is relativistic, going
approximately as $\Gamma^2$, where $\Gamma$ is the bulk Lorentz
factor. But the jet speeds required would be much larger than those
implied by sidedness analysis of 3C\,66B (H96) and in any case a
corresponding large increase in observed IC emission is only seen if
the jet is at small angles to the line of sight; this model cannot
work in 3C\,66B if the angle to the line of sight is $\sim 45^\circ$,
as estimated by Giovannini \etal\ (2001). The most promising seed
photon population is the light from a hidden BL Lac in the nucleus of
3C\,66B. Using the faster (and therefore more favourable) beaming
speed estimated by Giovannini \etal , $\beta = 0.99$, we estimate that
this photon population may come within an order of magnitude of
producing the required energy density in photons; this is close enough
that comparatively small departures from equipartition (a factor of a
few in $B$) can bring the predicted X-ray flux up to the observed
levels. A contribution to the X-ray flux of knot B from this process,
which could be significant if the beaming cone angle is less than the
jet opening angle (as it is for $\beta = 0.99$), might help to explain
the radio-X-ray offset seen for that feature, though some fine-tuning
(e.g.\ of magnetic field strength) would be necessary to ensure that
the IC emission did not dominate and cause knot B's spectrum to
resemble knot A's. However, our calculation is carried out assuming
that knot A is at rest with respect to the nucleus. If, as seems
likely, knot A is moving at relativistic speeds away from the nucleus,
the effectiveness of this process at producing observed X-rays is
reduced. In particular, if the jet is intrinsically two-sided and the
X-ray emission comes from inverse-Compton scattering of nuclear
photons, the jet to counterjet brightness contrast of knot A in the
radio and X-ray requires $\beta \ga 0.75$, which would push down the
effective photon energy density. Very roughly, we might expect this to
go down as $\Gamma^2$ (cf.\ Celotti \etal\ 2001), so factors of at
least 2 are probably involved, requiring correspondingly larger
departures from equipartition.

\section{Comparison with other sources}

\subsection{M87}

\begin{figure}
\epsfxsize 8.5cm
\epsfbox{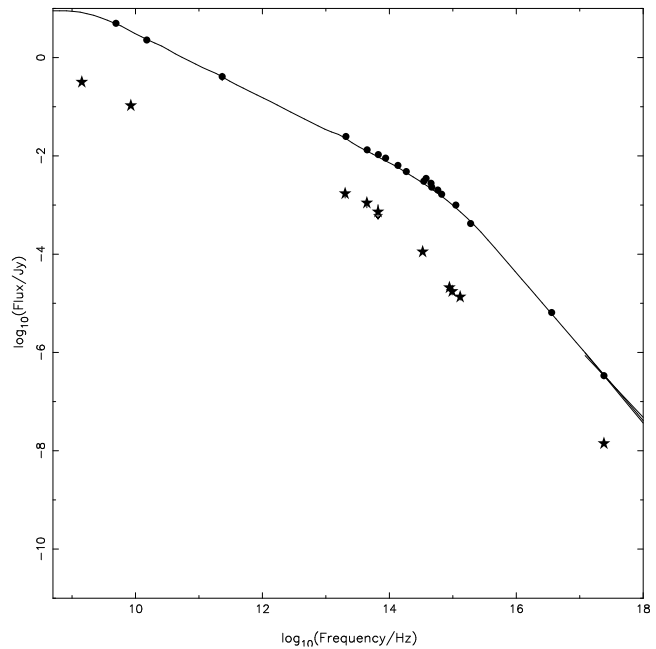}
\caption{The radio-to-X-ray spectrum of the ABC complex of M87. Data
points are taken from B\"ohringer \etal\ (2001), with the exception of
the EUVE point of Bergh\"ofer \etal (2000). The solid line shows a
broken power-law spectrum fitted to the data. The points lying below
the line, marked with stars, are the data for the whole jet of 3C\,66B
plotted in Fig.\ \ref{netspec}.}
\label{m87}
\end{figure}

The X-ray emission detected from the complex of knots A, B and C in
M87 with {\it Einstein}, {\it ROSAT} and {\it XMM-Newton} (Biretta,
Stern \& Harris 1991, Neumann \etal\ 1997, B\"ohringer \etal\ 2001)
seems likely to be synchrotron. The X-ray spectral index is found
by B\"ohringer \etal\ to be $1.45 \pm 0.1$, and the EUVE data point at
$4 \times 10^{16}$ Hz (Bergh\"ofer, Bowyer \& Korpela 2000) lies on
the line with $\alpha \approx 1.4$ connecting optical and X-ray data
points. This makes it unlikely that the jet spectrum can be fitted
with standard aged synchrotron models, as suggested by Perlman \etal\
(2001). M87's radio-to-X-ray jet spectrum is plotted in Fig.\
\ref{m87}, with comparison points from 3C\,66B (taken from Fig.\
\ref{netspec}); the overall spectra are strikingly alike, with similar
low-energy and high-energy spectral indices (the X-ray spectral
indices are consistent at the joint $1\sigma$ level if we assume the
simple power-law model for 3C\,66B's spectrum) and a similar spectral
break at around $10^{14}$ Hz. We infer that the unknown acceleration
process responsible for 3C\,66B's jet spectrum is also at work in M87.

The AB complex in M87 is in several ways analogous to knot B in
3C\,66B. They are at similar projected distances, $\sim 2$ kpc, from
their active nuclei, and they both represent the point where a narrow,
faint radio jet abruptly brightens. Because of this it is significant
that B\"ohringer \etal\ confirm that the X-ray peak of the ABC complex
in M87 is offset towards the core by $1.5 \pm 0.5$ arcsec, or $0.12
\pm 0.04$ projected kpc. This compares very well with the offset of
$0.21 \pm 0.03$ projected kpc we see in knot B of 3C\,66B.  The
similarity between the two sources would tend to support the model,
discussed above, in which the X-ray marks the location of a shock and
particle acceleration. It will be important to see whether this offset
can be seen in other sources and in other knots of M87 --- the model
predicts that the size of the offset depends quite strongly on the
magnetic field strength in the knot, which should easily be testable
with {\it Chandra} observations. It will also be interesting to
see whether there is any anomalous X-ray brightness of the inner jet
of M87, comparable to the situation in 3C\,66B's knot A.

\subsection{Centaurus A}

In Centaurus A, the jet X-rays have also been modelled as synchrotron
emission, and their spectrum is similar to that of the jet in 3C\,66B
(Turner \etal\ 1997), but no optical constraints are available. Kraft
\etal\ (2000) use {\it Chandra} HRC data to show that there is a good
detailed correspondence between individual knots in the radio and
X-ray regimes. We have examined archival {\it Chandra} ACIS-I data on
Centaurus A and confirm the result of Turner \etal\ that the X-ray
spectral index of the inner jet is $1.3 \pm 0.1$, very similar to the
values seen in 3C\,66B and M87. Intriguingly, the inner knot of Cen
A's X-ray jet [A1, in the notation of Burns, Feigelson \& Schreier
(1983)] has an X-ray-to-radio ratio ten times higher than its nearest
neighbours (A2, A3, A4) but a spectral index which is identical to
theirs within the errors; the similarity to the behaviour of 3C\,66B's
knot A is striking. The well-constrained steep spectrum of Cen A's
knot A1 and the similar distances of knots A1 and A2 from the nucleus
both suggest that an inverse-Compton model is unlikely to be viable in
explaining the X-ray excess in knot A1 in Cen A. New high-resolution
radio maps available to us show that there are also substantial
offsets between some of the radio and X-ray knots in the Cen A
jet. These data will be discussed in more detail elsewhere.

\subsection{3C\,273}

M87 and Cen A are both similar in total radio power to 3C\,66B. By
contrast, in the third pre-{\it Chandra} X-ray jet, that of the
luminous quasar 3C\,273, it has been clear for some time that a
synchrotron model may have difficulties (Harris \& Stern 1987, R\"oser
\etal\ 2000). More recently, two groups have analysed new {\it
Chandra} images (Sambruna \etal\ 2001, Marshall \etal\ 2001). Sambruna
\etal\ argue that X-ray flux densities from several regions in the jet
lie above a spectral extrapolation from the optical continuum, which
would imply that the X-rays cannot be synchrotron emission from the
population of electrons responsible for the radio and optical
jets. Instead, Sambruna \etal\ favour a model in which the jet is
highly relativistic and the X-rays are inverse-Compton scattered
microwave background photons. Marshall \etal\ interpret the optical
data rather differently and, despite finding rather flat X-ray
spectra ($\alpha \sim 0.7$), suggest that the emission can be modelled
satisfactorily as synchrotron, with a single power law connecting the
radio to the X-ray in the innermost knot and spectral steepening near
the optical for the others. In either case, 3C\,273 is probably not
directly comparable to 3C\,66B, but it is interesting that its
X-ray-to-radio ratio decreases with distance along the jet, which is
again suggestively similar to the differences we see in 3C\,66B
between knot A and the rest of the source.

\subsection{New jets}

The emission mechanisms are more uncertain for the jets newly
discovered with {\it Chandra}. In PKS 0637$-$752, the model that seems
to find most widespread favour is the boosted-inverse-Compton model
(Tavecchio \etal\ 2000), and such a model is also possible for Pictor
A (Wilson \etal\ 2001). However, the required extreme relativistic
motions on kpc scales are inconsistent with the much lower jet speeds
inferred in large samples of FRIIs from jet sidedness and prominence
statistics (Wardle \& Aaron 1997, Hardcastle \etal\ 1999). It does not
seem likely that this type of model will be viable in general for the
jets in FRIs, since much detailed study suggests that these are at
best mildly relativistic on kpc scales (e.g.\ Laing 1996; Laing \etal\
1999), and since it works best for jets close to the line of sight,
whereas the evidence seen so far with {\it Chandra} suggests that
essentially {\it every} FRI radio source with a bright radio jet also
has a detectable X-ray jet.

For the new FRI jets discovered by Worrall \etal\ (2001) less is known
about the radio-optical spectra, but in at least one of them, B2
0755+37, the level of the known optical jet (Capetti \etal\ 2000)
permits a synchrotron model for the X-rays with a spectrum very
similar to that of 3C\,66B.  We have data in hand for four other FRI
sources with X-ray jets, two of which have known optical jet
emission. At least two of them, like 3C\,66B, show evidence for an
anomalously X-ray bright inner jet.

\section{Conclusions}

We have detected a new X-ray jet in the FRI radio galaxy 3C\,66B. The
{\it Chandra} ACIS-S image shows a good overall correspondence between
X-ray and radio knots, although more detailed analysis shows
significant variation in the X-ray-to-radio ratio along the jet, with
the inner knot showing anomalously bright X-ray emission for its radio
strength. The brightest radio and X-ray knot, knot B, is significantly
closer to the core in the X-ray images than on the radio map, which
may be evidence that the knots are privileged sites for high-energy
particle acceleration.

3C\,66B's X-ray jet seems likely to be due to synchrotron emission,
although it is possible that some of the X-ray brightness of the inner
regions of the jet is due to inverse-Compton scattering of light from
the BL Lac nucleus. A synchrotron explanation is possible for all the
X-ray jets in other low-power, FRI sources known to us, if the
radio-optical-X-ray synchrotron spectra are similar, with a flat
($\alpha \approx 0.5$) radio spectrum breaking to a steeper values
($\alpha \approx 1.3$) in the infra-red or optical. However,
multi-frequency optical observations and good X-ray spectroscopy are
required to test this conclusion. An X-ray jet with a strong upper
limit on optical jet flux would pose a challenge to this simple
picture. We can also expect to see a more complex situation when {\it
Chandra} images of M87 become available.

If the X-ray emission in 3C\,66B and other FRI sources is synchrotron,
there is almost no way to avoid the necessity of {\it in situ}
particle acceleration throughout the X-ray-emitting regions of FRI
jets. The radio-optical-X-ray spectrum of 3C\,66B and the spatial
distribution of the optical and X-ray emission are hard to explain in
any standard particle acceleration/ageing model. Processes such as
magnetic field reconnection (e.g.\ Birk \& Lesch 2000) may need to be
incorporated into the standard model of FRIs if {\it Chandra}
continues to find large numbers of X-ray synchrotron jets in these
objects.

\bsp
\clearpage
\end{document}